\documentclass[12pt]{article}

\begin{document}

\title{{\textbf{SU(3) MIXING FOR EXCITED MESONS}}}
\author{W.S. Carvalho$^{\S }$\thanks{
weuber@acd.ufrj.br} , A.S. de Castro$^{\S \S }$\thanks{
castro@feg.unesp.br} \ and A.C.B. Antunes$^{\S \S \S }$\thanks{
antunes@if.ufrj.br} \\
\\
$^{\S }$Instituto de Geoci\^{e}ncias\\
Universidade Federal do Rio de Janeiro\\
21949-000 Rio de Janeiro RJ - Brasil\\
\\
$^{\S \S }$ UNESP/Campus de Guaratinguet\'{a} - DFQ\\
C.P. 205\\
12516-410 Guaratinguet\'{a} SP - Brasil\\
\\
$^{\S \S \S }$ Instituto de F\'{i}sica\\
Universidade Federal do Rio de Janeiro\\
C.P. 68528\\
21945-970 Rio de Janeiro RJ - Brasil}
\date{}
\maketitle

\begin{abstract}
The SU(3)-flavor symmetry breaking and the quark-antiquark annihilation
mechanism are taken into account for describing the singlet-octet mixing for
several nonets assigned by Particle Data Group(PDG). This task is approached with 
the mass matrix formalism.
\end{abstract}

\newpage

\section{Introduction}

In the constituent quark model the mesons are considered as bound states of
a quark and an antiquark. Taking into account the SU(3)-flavor symmetry the
mesons are either in an SU(3) singlets or octets: \textbf{3}$\otimes $%
\textbf{3=1}$\oplus $\textbf{8}. Nonetheless, due to the SU(3)-symmetry
breaking the isoscalar physical states appear as mixtures of the singlet and
octet members. This singlet-octet mixing is also called SU(3) mixing. The
inability of the Gell-Mann-Okubo mass formula \cite{GMO} to adjust the
masses of the pseudoscalar mesons has been considered as a suggestion for
the inclusion of other effects such as the quark-antiquark annihilation into
gluons. The fail of an SU(3)-invariant annihilation amplitude in attempting
to solve the $\eta $-$\eta ^{\prime }$ mass splitting \cite{schw}, \cite
{close} led De Rujula \textit{et al}. \cite{ruju} to propose that the
quark-antiquark annihilation mechanism might not be SU(3)-invariant.

In a previous paper \cite{w1} the $\eta $-$\eta ^{\prime }$ mass splitting
was explained in a SU(3)-symmetry breaking framework. The physical states
are mixtures of the isoscalar singlet and octet states and the amplitudes of
quark-antiquark annihilation into gluons as well as the binding energies are
supposed to be flavor dependent. Within this formulation an extended
expression for Schwinger's sum rule is satisfied. Also the SU(3) mixing
angle obtained, $\theta =-19.51^{o}$, is consistent with the experimental
data $(\theta \simeq -20^{o})$ from $\eta $ and $\eta ^{\prime }$ decays
into pions \cite{PDG1}. As a very natural extension of the previous paper we
assume the SU(2)-symmetry breaking in the SU(3) mixing framework \cite{w2}.
In this way the pseudoscalar neutral mesons are mixtures of isoscalar and
isovector states and the neutral pion takes part in the mixing scheme.
This model works well, but the result gives a hint that some significant
effect possibly has not been considered. The strange result is that the
ratio $m_{s}/m_{u}\simeq 2$ takes a somewhat large value, in comparison with
those  used in the constituent quark models ($m_{s}/m_{u}\simeq
1.3\ldots 1.8$). Our formulation is incompatible with fundamental models. If
current quark masses were used the free parameters of the model would not be
able to fit the masses of $\eta $ and $\eta ^{\prime }$. In addition, the
correct singlet-octet mixing angle would not be obtained.

The $\eta $-$\eta ^{\prime }$ mixing scheme could be enlarged by the
inclusion of gluonic degrees of freedom. The $\iota (1440)$ was interpreted
as a strong glueball candidate due to its enhanced production in a
gluon-rich channel \cite{scharre}. The $\iota (1440)$, with the same quantum
numbers as the $\eta $ and $\eta ^{\prime }$ system, motivated the study of
the $\eta $-$\eta ^{\prime }$-$\iota $ mixing scheme \cite{japa}-\cite
{jousset}. Recently, the mass region near to $\iota (1440)$ has been
resolved into two states $\eta ^{\prime \prime }(1410)$ and $\eta (1490)$ 
\cite{bai}. The first one has been interpreted as being mainly a glueball
mixed with $q\bar{q}$ and the second one as mainly a $s\bar{s}$ radially
excited state \cite{kitamura},\cite{closea}. Therefore one is tempted to
identify $\eta ^{\prime \prime }(1410)$ as the remaining physical state in
this extended mixing scheme \cite{kitamura}-\cite{pred} for ground states.
On the other hand, the state $\eta (1490)$ is interpreted as a partner of
the radially excited state $\eta (1295)$ \cite{closea}. The states $\eta
(1295)$ and $\eta (1490)$ are the physical manifestations of mixtures among
2S excited states including solely light and strange quarks \cite{kitamura}.
In a recent paper \cite{w3} we describe the $\eta $-$\eta ^{\prime }$-$\eta
^{\prime \prime }$ and $\eta (1295)$-$\eta (1490)$ systems with the same
formalism used in Ref. \cite{w1} but enlarging the mixing scheme to include
glueballs. The small overlapping of the respective mass intervals suggests
the possibility of mixing among ground states and radial excitations as
considered by \cite{lipkinb}, however, in a first approximation, we assume
that this 1S-2S mixing may be neglected. In searching for the best results
of the branching ratios and of the decay widths involving the $\eta $, $\eta
^{\prime }$ and $\eta ^{\prime \prime }$ mesons we have fixed all the
parameters of the problem. This enlarged mixing scheme furnishes
satisfactory results for the experimental data and improves the high value
for the ratio $m_{s}/m_{u}$ obtained in Ref. \cite{w3}. We obtained $%
m_{s}/m_{u}=1.772$. Finally we extend the mixing scheme to the excited
states using the value of $m_{s}/m_{u}$ determined for the ground state.

The nonet of axial ($1^{++},1^{3}P_{1}$) and tensor ($2^{++},1^{3}P_{2}$)
mesons are well established \cite{PDG2}. The axial nonet consists of the
isodoublet $K_{1A}(1340)$, the isovector $a_{1}(1260)$ and the isoscalars $%
f_{1}(1285)$ and $f_{1}(1510)$. The $K_{1A}$ is a mixture of $K_{1}(1270)$
and $K_{1}(1470)$ with a close to 45$^{\circ }$ mixing angle \cite{k1}$.$
The tensor nonet is formed by the isodoublet $K_{2}^{*}(1430)$, the
isovector $a_{2}(1320)$ and the isoscalars $f_{2}(1270)$ and $f_{2}^{\prime
}(1525)$. Nonetheless, there are extra isoscalar states with quantum numbers
and masses permitting that they can be interpreted as partners of the nonets
of axial and tensor mesons. The axial state $f_{1}(1420)$, observed in two
experiments \cite{Gavillet}, has been considered by some authors \cite
{Bityukov} as a possible candidate to exotic. On the other side, there are
two candidate to exotic tensor states: $f_{2}(1640)$ \cite{Alde} and $%
f_{J}(1710)$ \cite{Augustin}. There is a controversy about the value of the
spin of the $f_{J}(1710)$: it may be a scalar or a tensor state \cite{bes}.
In other paper \cite{w4} we approached the problem of axial and tensor
mesons where the candidates to exotics $f_{1}(1420)$ and $f_{2}(1640),$ or $%
f_{2}(1710)$, are supposed to be components of a quarkonia-gluonia mixing
scheme similar to that previously applied to the pseudoscalar mesons \cite
{w1}. In this last paper $m_{s}/m_{u}=1.772$ determined in Ref. \cite{w3}
has been used as an input. The predictions of the model for branching ratios
and electromagnetic decays are incompatible with the experimental results.
These facts suggest the absence of gluonic components in the axial and
tensor isosinglet mesons analyzed. On the other hand, the interpretations of
the states $f_{1}(1420)$, $f_{1}(1510)$, $f_{2}(1640)$ and $f_{J}(1710)$ are
controversial and, moreover, some of them need confirmation. The same mixing
scheme was not applied to the scalar states because only the assignment for
the scalar isodoublet is well-established.

Here we analyze the mixing scheme for the nonets listed in Table 13.2 of
Particle Data Group (PDG) \cite{PDG3} which have all the members suggested,
including the scalar states and excepting the lowest pseudoscalar states ($%
\pi $, $K$, $\eta $, $\eta ^{\prime }$). For all intents and purposes we
ignore any quarkonia-gluonia interference. We also assume the SU(2)
invariance which is justified by a preceding work \cite{w2} in which we have
shown that the SU(2)-symmetry breaking is important to the mass splitting
between the $\pi ^{0}$ and $\pi ^{\pm }$, but it has negligible effects in
the $\eta $-$\eta ^{\prime }$ mixing. We will suppose that the isospin
symmetry breaking causes no mixing between the isoscalar members of the
excited nonets.

\section{The mass matrix formalism}

Several kinds of the mixing schemes has been proposed to give account of the
peculiar properties of the isoscalar mesons. In some schemes the physical
states are written as linear combinations of pure quarkonia and gluonia
states. The linear coefficients are generally related to the rotation angles
and may be determined by the decay properties of, or into, the physical
mesons \cite{antunes}, \cite{jousset}, \cite{closea}, \cite{pred}, \cite
{harber}, \cite{fclose}. Another approach, in which the interference is
considered at a more fundamental level, consists in writing a mass matrix
for the physical states in the basis of the pure quarkonia and gluonia
states. The elements of this mass matrix are obtained from a model that
describes the process of interference. The mixtures of the basic states are
induced by the off-diagonal elements. Thus, these elements must contain the
amplitudes for transitions from one to another states of the basis. The
eigenvalues of that matrix give the masses of the physical states and the
corresponding eigenvectors give the proportion of quarkonia and gluonia in
each meson \cite{rosner}, \cite{kitamura}, \cite{fritzch1}.

In Ref. \cite{w1}, \cite{w2}, \cite{w3}, \cite{w4} we have adopted a mixing
scheme based on a mass matrix approach. The flavor-dependent annihilation
amplitudes and binding energies are the responsible mechanisms for the
quarkonia-gluonia mixing. Here a brief review of the mass matrix formalism
we have used in previous papers is outlined only for the quarkonia mixing.
The mass matrix in the basis $|u\bar{u}>$, $|d\bar{d}>$ and $|s\bar{s}>$,
including flavor-dependent binding energies and annihilation amplitudes, has
matrix elements given by 
\begin{equation}
\mathcal{M}_{ij}=(2m_{i}+E_{ij})\delta _{ij}+A_{ij}  \label{eq11}
\end{equation}
where $i,j=u,d,s$. The contribution to the elements of the mass matrix are:
the rest masses of the quarks $m_{i}$, the eigenvalues $E_{ij}$ of the
Hamiltonian for the stationary bound state ($ij$) and the amplitudes $A_{ij}$%
, that account for the possibility of quarkonia-gluonia transitions. As in
the previous papers we assume that $E_{ij}$ and $A_{ij}$ are not
SU(3)-invariant quantities. Another basis also used consists of the
isoscalar singlet and octet of the SU(3)

\begin{eqnarray}
|1>\; &=&\frac{1}{\sqrt{3}}\left( \sqrt{2}|N>+\;|S>\right)  \label{s} \\
&&  \nonumber \\
|8>\; &=&\frac{1}{\sqrt{6}}\left( \sqrt{2}|N>-\;2|S>\right)  \label{o}
\end{eqnarray}

\noindent where this basis is written in a form that presents a segregation
of strange and nonstrange quarks,

\begin{eqnarray}
|N>\;&=&\frac{1}{\sqrt{2}}\left( |u\bar{u}>+\;|d\bar{d}>\right) \\
&&  \nonumber \\
|S>\;&=&|s\bar{s}>
\end{eqnarray}

\noindent

\noindent Besides these states we need also the isovector states

\begin{equation}
|\tilde{\pi}^{0}>\;=\frac{1}{\sqrt{2}}(|u\bar{u}>-\;|d\bar{d}>)  \label{p}
\end{equation}

\noindent In this basis the mixing among the isoscalar and isovector states
is caused by isospin symmetry breaking terms. Therefore, assuming the exact
SU(2)-flavor symmetry, one needs only consider the subspace spanned by the
isoscalar states when the mass matrix reduces to a 2x2 matrix $\mathcal{M}%
_{0}:$

\begin{equation}
\mathcal{M}_{0}=\left( 
\begin{array}{cc}
m_{8} & m_{18} \\ 
m_{18} & m_{1}
\end{array}
\right)  \label{m0}
\end{equation}

\noindent where

\begin{eqnarray}
m1 &=&\frac{2}{3}\left( 2m_{u}+m_{s}\right) +\frac{1}{3}\left(
2E_{uu}+E_{ss}\right) +A_{11}  \label{m1} \\
&&  \nonumber \\
m_{8} &=&\frac{2}{3}\left( m_{u}+2m_{s}\right) +\frac{1}{3}\left(
E_{uu}+2E_{ss}\right) +A_{88}  \label{m8} \\
&&  \nonumber \\
m_{18} &=&\frac{2\sqrt{2}}{3}\left( m_{u}-m_{s}\right) +\frac{\sqrt{2}}{3}%
\left( E_{uu}-E_{ss}\right) +A_{18}  \label{m18}
\end{eqnarray}

\noindent and

\begin{eqnarray}
A_{88} &=&\frac{2}{3}\left( A_{uu}-2A_{us}+A_{ss}\right)  \label{a88} \\
&&  \nonumber \\
A_{11} &=&\frac{1}{3}\left( 4A_{uu}+4A_{us}+A_{ss}\right)  \label{a11} \\
&&  \nonumber \\
A_{18} &=&\frac{\sqrt{2}}{3}\left( 2A_{uu}-A_{us}-A_{ss}\right)  \label{a18}
\end{eqnarray}

\noindent Using the mass relations for the isovector and isodoublet members,

\begin{eqnarray}
M_{1} &=&2m_{u}+E_{uu}  \label{M1} \\
&&  \nonumber \\
M_{1/2} &=&m_{u}+m_{s}+E_{us}  \label{M1/2}
\end{eqnarray}

\noindent where the annihilation effects are absent, only the rest masses of
the quarks and the binding energies contribute to the physical masses. The
notation uses subscripts in $M$ to identify the isospin. Defining

\begin{equation}
M_{1/2}^{(\varepsilon )}=M_{1/2}+\varepsilon  \label{m11}
\end{equation}

\noindent where

\begin{equation}
\varepsilon =\frac{E_{uu}+E_{ss}}{2}-E_{us}  \label{eps}
\end{equation}

\noindent the elements of the mass matrix $\mathcal{M}_{0}$ are found to be

\begin{eqnarray}
m_{1} &=&\frac{1}{3}\left( 2M_{1/2}^{(\varepsilon )}+M_{1}\right) +A_{11}
\label{m111} \\
&&  \nonumber \\
m_{8} &=&\frac{1}{3}\left( 4M_{1/2}^{(\varepsilon )}-M_{1}\right) +A_{88}
\label{m1188} \\
&&  \nonumber \\
m_{18} &=&\frac{2\sqrt{2}}{3}\left( M_{1}-M_{1/2}^{(\varepsilon )}\right)
+A_{18}  \label{m188}
\end{eqnarray}

\noindent The above results show that the SU(3)-symmetry breaking gives rise
to off-diagonal elements in the mass matrix. These elements are generated
not only by the gluon annihilation amplitudes but also by influences due to
the differences in the binding energies. These off-diagonal elements are
responsible for the mixing effects among the states composing the physical
mesons. We adopt an expression for the amplitude of the process $q\bar{q}%
\leftrightarrow gg\leftrightarrow q^{\prime }\bar{q}^{\prime }$ similar to
that of Cohen \textit{et al.} \cite{lipkina} and Isgur \cite{isgur2}, where
the numerator of the two-gluon annihilation amplitude expression is assumed
to be a SU(3)-invariant parameter, which means that we parameterize the
annihilation amplitude in the form \vspace{0.2cm}

\begin{equation}
A_{qq^{\prime }}=\frac{\Lambda }{m_{q}m_{q^{\prime }}}  \label{a}
\end{equation}

\vspace{0.3cm} \noindent The phenomenological parameter $\Lambda $ is to be
determined. Then, the amplitudes become

\begin{eqnarray}
A_{11} &=&\frac{1}{2}(2+r_{1})^{2}r_{2}  \label{a111} \\
&&  \nonumber \\
A_{88} &=&\frac{2}{3}(1-r_{1})^{2}r_{2}  \label{a188} \\
&&  \nonumber \\
A_{18} &=&\frac{\sqrt{2}}{3}\left( 2+r_{1}\right) \left( 1-r_{1}\right) r_{2}
\label{a118}
\end{eqnarray}

\noindent where

\begin{eqnarray}
\frac{1}{r_{1}} &=&\frac{m_{s}}{m_{u}}  \label{r1} \\
&&  \nonumber \\
r_{2} &=&\frac{\Lambda }{m_{u}^{2}}  \label{r2}
\end{eqnarray}
\noindent The invariants of the mass matrix $\mathcal{M}_{0}$ under a
unitary transformation give the following mass relations for the isoscalar
physical states:

\begin{eqnarray}
M+\widetilde{M} &=&\rm{tr }(\mathcal{M}_{0})  \label{ttr} \\
&&  \nonumber \\
M\times \widetilde{M} &=&\rm{det }(\mathcal{M}_{0})  \label{ddet}
\end{eqnarray}

\noindent where $M$ and $\widetilde{M}$ are the eigenvalues of the mass
matrix $\mathcal{M}_{0}$ (masses of the isoscalar physical states). Their
corresponding eigenvectors are the physical states $|M>$ and $|\widetilde{M}%
> $ which are mixtures of $|1>$ and $|8>$:

\begin{eqnarray}
|M>\; &=&\cos (\theta )\;|8>-\;\sin (\theta )\;|{1}>  \label{eta} \\
&&  \nonumber \\
|\widetilde{M}>\; &=&\sin (\theta )\;|{8}>+\;\cos (\theta )\;|{1}>
\label{etal}
\end{eqnarray}

\noindent where the coefficients of the eigenvectors are written in terms of
the singlet-octet mixing angle given by

\begin{equation}
\theta =\arctan \left( \frac{m_{8}-M}{m_{18}}\right)  \label{teta}
\end{equation}

\noindent In terms of strange and nonstrange quarks (\ref{eta})-(\ref{etal})
can be written as

\begin{eqnarray}
|M &>&\;=X|N>+Y|S>  \label{nova1} \\
&&  \nonumber \\
|\widetilde{M} &>&\;=\widetilde{X}|N>+\widetilde{Y}|S>  \label{nova2}
\end{eqnarray}

\noindent where

\begin{equation}
X\;=\widetilde{Y}=\frac{\cos (\theta )-\sqrt{2}\sin (\theta )}{\sqrt{3}}%
,\qquad Y=-\widetilde{X}=-\frac{\sqrt{2}\cos (\theta )+\sin (\theta )}{\sqrt{%
3}}  \label{x2}
\end{equation}

\vspace{0.2cm}

\noindent Eliminating $A_{11}$ from (\ref{ttr}) and (\ref{ddet}) we obtain
the generalized Schwinger sum rule:

\begin{equation}
(M+\widetilde{M})(4M_{1/2}^{(\varepsilon )}-M_{1})-3M\widetilde{M}=4\left[
2M_{1/2}^{(\varepsilon )}-(1-r_{1}^{2})r_{2}\right] (M_{1/2}^{(\varepsilon
)}-M_{1})+3M_{1}^{2}  \label{schw}
\end{equation}

\vspace{0.2cm}

\noindent \noindent To our knowledge this generalized sum rule was obtained
for the first time in Ref. \cite{w1}. Note that the ordinary Schwinger sum
rule \cite{schw} can be recovered doing $r_{1}=1$ in (\ref{schw}). Equations
(\ref{ttr}) and (\ref{ddet}) can also be solved for $r_{1}$ and $r_{2}$
giving

\begin{eqnarray}
\frac{m_{s}}{m_{u}} &=&\frac{\sqrt{2}}{2}\sqrt{\frac{\left( M-M_{1}\right)
\left( \widetilde{M}-M_{1}\right) }{\left( \widetilde{M}+M_{1}-2M_{1/2}^{(%
\varepsilon )}\right) \left( 2M_{1/2}^{(\varepsilon )}-M-M_{1}\right) }}
\label{r11} \\
&&  \nonumber \\
\frac{\Lambda }{m_{u}^{2}} &=&\frac{\left( \widetilde{M}-M_{1}\right) \left(
M-M_{1}\right) }{4\left( M_{1/2}^{(\varepsilon )}-M_{1}\right) }  \label{r22}
\end{eqnarray}

\noindent The invariants of the mass matrix are functions of $m_{s}/m_{u}$, $%
\Lambda /m_{u}^{2}$ and $\varepsilon $. These quantities are not all free.
The equations (\ref{ttr}) and (\ref{ddet}) impose some constraints among
them. The equations are to be solved for $\Lambda /m_{u}^{2}$ and $%
\varepsilon $ by considering $m_{s}/m_{u}$ in a range of values consistent
with those usually adopted when using constituent quark masses in
nonrelativistic quark model ($m_{s}/m_{u}=1.3\ldots 1.8$). For finding the
solutions one needs to solve a second degree algebraic equation. One of
those solutions is an extraneous root and the criterion to get rid of it is
the comparison with the solution obtained for the SU(3) mixing angle (\ref
{teta}) in the case of SU(3)-invariant amplitudes and binding energies. Our
choice consists in the mixing angle nearest to that SU(3)-invariant mixing
angle.

\section{Mixing in excited states}

The mixing scheme briefly presented in the preceding section, ignoring any
quarkonia-gluonia mixing, is now applied to the excited members of the
nonets. The attention will be paid to the referred assignments in the Table
13.2 of the PDG \cite{PDG3}, even for the cases which are controversial.
These results, corresponding to the range $m_{s}/m_{u}=1.3\ldots 1.8,$ are
summarized in Table 1.

\subsection{$1$ $^{1}S_{0}\mathbf{\ (}0^{-+}\mathbf{)}$}

The ground-state pseudoscalar nonet ($\pi $, $K$, $\eta $, $\eta ^{\prime }$%
) has already been considered in Ref. \cite{w3}, where an enlarged mixing
scheme including gluonia has been shown to be necessary. Putting to test the
present mixing scheme for this nonet without gluonic degrees of freedom ends
in a complete fiasco in the range of $m_{s}/m_{u}$ considered.

\subsection{$1$ $^{3}S_{1}\mathbf{\ (}1^{--}\mathbf{)}$}

The ground-state vector nonet ($\rho ,K^{*}(892),\omega ,\phi $) is well
established since a long time ago. It presents a SU(3) mixing angle near to
ideal $\omega -\phi $. It can be found that $\phi $ presents $%
99.9\%\ldots100\%$ of strange quarks and mixing angles in the range $%
36.9^{o}\ldots36.4^{o}$. These values are to be compared with this one
listed by PDG ($\theta =36^{o}$).

\subsection{$1$ $^{1}P_{1}\mathbf{\ (}1^{+-}\mathbf{)}$}

We found that the content of strange quarks in $h_{1}(1380)$ is much higher
than in its isoscalar partner. This result is supported by the experimental
data which show $h_{1}(1380)\rightarrow KK^{*}(892)+c.c.$ and $%
h_{1}(1170)\rightarrow \rho \pi $ being the only ones decay modes seen, at
least up to now.

\subsection{$1$ $^{3}P_{0}\mathbf{\ (}0^{++}\mathbf{)}$}

For this nonet we found that $f_{0}(1370)$ presents $89.7\%_{+4.9\%}^{-18.5%
\%}\ldots 94.1\%_{+2.9\%}^{-10.9\%}$ of strange quarks and $\theta
=-73.4^{o}\ _{+5.3^{o}}^{-13.8^{o}}\ldots -68.8^{o}\ _{+3.9^{o}}^{-10.2^{o}}$
These values were found taking into account that the broad resonance $%
f_{0}(1370)$ has mass equal to ($1.35\pm 0.15)$ GeV. It is worthwhile to
remark that among the two candidates for the $I=1$ ($a_{0}(980),a_{0}(1450)$%
) states and the four ones for $I=0$ ($%
f_{0}(400-1200),f_{0}(980),f_{0}(1370),f_{0}(1710)$) acceptable results were
found only for the isovector $a_{0}(1450)$ and for the isoscalars $%
f_{0}(1370)$ and $f_{0}(1710)$, namely the states listed in Table 13.2 of
PDG. It should be highlighted, though, that $f_{0}(1710)$\ contains only a
small fraction of strange quarks in contrast to the indication of the PDG
based on the naive quark model. In addition, it is observed that $%
f_{0}(1710) $ has a dominant $K\overline{K}$ decay mode and $f_{0}(1370)$
couples more strongly to $\pi \pi $ than to $K\overline{K}$.

\subsection{$1$ $^{3}P_{1}\mathbf{\ (}1^{++}\mathbf{)}$}

The $f_{1}(1420)$ competes for a $s\overline{s}$ assignment with percentages
of $94.6\%\ldots 97.7\%$ and mixing angles in the range $-41.3^{o}\ldots
-45.9^{o}$  roughly agreement with $75\%\ldots 84\%$ and $\theta $ $\sim
-40^{o}$ obtained by Close\textit{\ et al.} \cite{fclose}. More recently Li 
\textit{et al. }\cite{li} obtained $92\%$ of $s\overline{s}$ in $f_{1}(1420)$
and $\theta $ $=-38.5^{o}$. As a matter of fact, they obtained $\sim 50^{o}$
and $51.5^{o}$, respectively, because they changed $|M>$ by $|\widetilde{M}>$%
, and vice versa, in (\ref{eta})-(\ref{etal}). The ratio of $J/\psi $
radiative branching ratios into $f_{1}(1285)$ and $f_{1}(1420)$ and the
ratio of the two-photon width of $f_{1}(1285)$ and $f_{1}(1420)$ are, using
the formulas in Ref. \cite{asei}, given by:

\begin{equation}
\frac{\Gamma _{\gamma \gamma }(\widetilde{f}_{1})}{\Gamma _{\gamma \gamma
}(f_{1})}=\left( \frac{5\widetilde{X}+\sqrt{2}\widetilde{Y}}{5X+\sqrt{2}Y}%
\right) ^{2}\left( \frac{\widetilde{M}}{M}\right) ^{3}  \label{R1}
\end{equation}

\begin{equation}
\frac{\Gamma _{\gamma \gamma }(\widetilde{f}_{1})}{\Gamma _{\gamma \gamma
^{*}}(f_{1})}=\left( \frac{5\widetilde{X}+\sqrt{2}\widetilde{Y}}{5X+\sqrt{2}Y%
}\right) ^{2}\left( \frac{\widetilde{M}}{M}\right) ^{3}  \label{R2}
\end{equation}

\begin{equation}
\frac{B(J/\psi \rightarrow \gamma \widetilde{f}_{1})}{B(J/\psi \rightarrow
\gamma f_{1})}=\left( \frac{\sqrt{2}\widetilde{X}+\widetilde{Y}}{\sqrt{2}X+Y}%
\right) ^{2}\left( \frac{\widetilde{P}}{P}\right) ^{3}  \label{R3}
\end{equation}

\begin{equation}
\frac{B(f_{1}\rightarrow \gamma \phi )}{B(f_{1}\rightarrow \gamma \rho )}=%
\frac{4}{9}\left( \frac{P_{\phi }}{P_{\rho }}\right) ^{3}\left( \frac{X}{Y}%
\right) ^{2}  \label{R4}
\end{equation}

\noindent where $f_{1}$ and $\widetilde{f}_{1}$ stand for $f_{1}(1285)$ and $%
f_{1}(1420)$, respectively. Our results are summarized in Table 2. In the 
table one can see that the ratio of (\ref{R1}) and (\ref{R3}) and (\ref{R2})
and (\ref{R3}) yield $0.39\ldots 0.36$. On the experimental side these
ratios yield $1.03\pm 0.92$ (an inferior limit) and $0.46\pm 0.40$,
respectively.

\subsection{$1$ $^{3}P_{2}\mathbf{\ (}2^{++}\mathbf{)}$}

For this nonet we found mixing angles in the range $30.1^{o}\ldots 31.5^{o}$
which are to be compared with the value $26^{o}$ presented by PDG and $%
27.5^{o}$ found by Li \textit{et al}. \cite{li2}. The ratio of branching
ratios, where $f_{2}$ and $\widetilde{f}_{2}$ stand for $f_{2}^{\prime
}(1525)$ and $f_{2}(1270)$, respectively, are given by

\begin{equation}
\frac{B(f_{2}\rightarrow \pi \pi )}{B(f_{2}\rightarrow K\overline{K})}=\frac{%
3X^{2}}{\left( \sqrt{2}Y+X\right) ^{2}}\left( \frac{{P_{\pi }}}{{P_{K}}}%
\right) ^{5}  \label{r6}
\end{equation}

\noindent

\begin{equation}
\frac{B(\widetilde{f}_{2}\rightarrow K\overline{K})}{B(\widetilde{f}%
_{2}\rightarrow \pi \pi )}=\frac{\left( \sqrt{2}\widetilde{Y}+\widetilde{X}%
\right) ^{2}}{3\widetilde{X}^{2}}\left( \frac{{P_{\pi }}}{{P_{K}}}\right)
^{5}  \label{r7}
\end{equation}

\noindent

\begin{equation}
\frac{B(J/\psi \rightarrow \gamma f_{2})}{B(J/\psi \rightarrow \gamma 
\widetilde{f}_{2})}=\left( \frac{\sqrt{2}X+Y}{\sqrt{2}\widetilde{X}+%
\widetilde{Y}}\right) ^{2}\left( \frac{P}{\widetilde{P}}\right) ^{3}
\label{r8}
\end{equation}

\noindent Our results and their comparison with the experimental data for
this nonet are summarized in Table 3.

\subsection{$1$ $^{1}D_{2}\mathbf{\ }(2^{-+}\mathbf{)}$}

We obtained values consistent with a near to ideal $\eta _{2}(1645)-\eta
_{2}(1870)$ mixing and the second isoscalar being dominantly composed of $s%
\overline{s}$ as speculated by PDG, although there are some expectations
that it may be an hybrid \cite{feclose}, \cite{ado}.

\subsection{$1$ $^{3}D_{3}\mathbf{\ }(3^{--}\mathbf{)}$}

For this nonet we found mixing angles in the range $31.4^{o}\ldots 32.4^{o}$
which are to be compared with the value $28^{o}$ presented by PDG.

\subsection{$1$ $^{3}F_{4}\mathbf{\ (}4^{++}\mathbf{)}$}

We found that $f_{4}(2220)$ is mainly a $s\overline{s}$ state. This result
agrees with the suggestion of PDG and had already been conjectured by
Godfrey \textit{et al.} \cite{go} and Blundell \textit{et al.} \cite{blu}.

\subsection{$2$ $^{1}S_{0}\mathbf{\ (}0^{-+}\mathbf{)}$}

For the first radial excitation of the pseudoscalar nonet we found that $%
\eta (1440)$ and $\eta (1295)$ presents almost an ideal mixing with the
first isoscalar being a $s\overline{s}$ state. Nevertheless, the $\eta
(1440) $ is now considered to be composed of two resonances: $\eta (1410)$
and $\eta (1490)$ \cite{bai}. The first one has been interpreted as being
mostly a glueball mixed with $q\bar{q}$ and the second one as mostly a $s%
\bar{s}$ radially excited state \cite{kitamura},\cite{closea}. The $\eta
(1410)$ has been identified as the remaining physical state in the
quarkonia-gluonia mixing scheme for the pseudoscalar ground states \cite
{kitamura}-\cite{w3}. On the other hand, the state $\eta (1490)$ is
interpreted as a partner of the radially excited state $\eta (1295)$ \cite
{kitamura}, \cite{closea}, \cite{w3}. With this point of view we found that $%
\eta (1490)$ is a $\sim 100\%$ $s\overline{s}$ state and the mixing angle is
in the range $-55.4^{o}\ldots -55.2^{o}$.

\subsection{$2$ $^{3}S_{1}\mathbf{\ (}1^{--}\mathbf{)}$}

The PDG proposes the $\rho (1450)$ to be the isovector partner for this
nonet, however we were unable to find consistent results even for the
candidate $\rho (1700)$. On the other hand, the state $\rho (1300)$ reported
by the LASS detector team \cite{lass}, without any entry in the PDG tables,
leads to results almost satisfactory. We found that $\phi (1680)$ has a
sizeable $s\overline{s}$ component ($89.7\%\ldots 96.1\%$), but is the $%
\omega (1420)$ which is mostly octet. This last result is in accord to the
experimental data which show that $\phi (1680)\rightarrow KK^{*}(892)+c.c.$
is the dominant decay for $\phi (1680)$ and besides $\omega (1420)$ has no
decay to $K\overline{K}$. It is worthwhile to note that  is the isoscalar $%
\omega (1420)$ which is mostly octet instead of the $\phi (1680)$ state. The
PDG suggests that the isodoublet $K^{*}(1410)$ could be replaced by the $%
K^{*}(1680)$ in this nonet. Unfortunately, with this replacement we are led
to unsatisfactory results for all the $\rho $ candidates.

\section{Conclusion}

In this paper we have shown that a mixing flavor approach similar to that 
used to describe the isosinglet states of the pseudoscalar meson nonet 
\cite{w1} can also be used also to describe isosinglet states for several
angular momentum and radially excited nonets. In this approach we assumed
SU(2) invariance. Moreover, we assumed that the constituent masses of the
quarks, the binding energies of the states and the gluon annihilation
amplitudes are not SU(3)-invariant quantities. The gluon annihilation
amplitudes were parameterized according to the prescriptions of Cohen 
\textit{et al.} \cite{lipkina} and Isgur \cite{isgur2}. In addition to these
assumptions we disregarded the presence of gluonic components in the
physical states. A linear 2x2 matrix formulation based in these assumptions
was applied to seven orbitally excited nonets and two radially excited
S-wave nonets.

The mixing scheme used in this paper works properly for the majority of the
isoscalar states listed in the Table 13.2 of the PDG \cite{PDG3}. Ten nonets
were analyzed and eight of them appear to be compatible with the
experimental predictions for their quark-antiquark contents, branching
ratios and radiative decays. Only in two cases our results mismatch the
experimental data. In these two cases the isoscalar states are not well
established. In the scalar sector there are many resonances competing to be
the isoscalar partners of this nonet. The mixing scheme only works using $%
a_{0}(1450)$, $f_{0}(1370)$ and $f_{0}(1710)$, the states listed in Table
13.2 of PDG, nevertheless we found unsatisfactory results. The current
status of the scalar nonet exclude any possibility to achieve a reliable
conclusion. For the 2$^{3}S_{1}$ sector a consistent result was reached
using the $\rho (1300)$, contrasting with the candidates listed by the PDG ($%
\rho (1450)$ and $\rho (1700)$). This point might be considered as a fail of
our mixing scheme but the existence of two $\rho $ states and maybe a third
one ($\rho (1300)$) would suggest a non-trivial interpretation for this
nonet.

To summarize, almost every nonet analyzed in this paper can be
satisfactorily described by our mixing scheme without any non-quark mesons.
The relative success of this approach suggests that it might be used as a
guide to the analyses of quark-antiquark contents of the physical mesons
participating of a specific nonet.

\bigskip

\bigskip

\noindent \textbf{Acknowledgments}

This work was partially supported by CNPq, FAPESP and FINEP.

\pagebreak

\newpage \pagebreak

\begin{table}[tbp]
\caption{SU(3) mixing angles for excited nonets. As done by PDG \protect\cite
{PDG3} {\protect\underline{the isosinglets mostly octet are listed first}}
and their percentual contents of strange quarks are also shown. The values
presented for $|S>$ and $\theta $ correspond to the range $%
m_{s}/m_{u}=1.3\ldots 1.8$. The values for the 1 $^{3}P_{0}$ nonet are found
taking into account the central value for the mass of $f_{0}(1370)$.}
\label{t1}
\begin{center}
{\small 
\begin{tabular}{|c|c|c|c|c|}
\hline\hline
N $^{2s+1}L_{J}$ & J$^{PC}$ & Nonet members & $|S>$ & $\theta $ \\ \hline
1 $^{3}S_{1}$ & $1^{--}$ & $\rho, K^{*}(892),$ \underline{$\phi$}$, \omega$
& $99.9\%\ldots 100\%$ & $36.9^{o}\ldots 36.4^{o}$ \\ 
1 $^{1}P_{1}$ & $1^{+-}$ & $b_{1}(1235), K_{1B},$ \underline{$h_{1}(1380)$}, 
$h_{1}(1170)$ & $98.0\%\ldots 98.9\%$ & $-62.9^{o}\ldots {-60.8^{o}}$ \\ 
1 $^{3}P_{0}$ & $0^{++}$ & $a_{0}(1450), K^{*}_{0}(1430),$ \underline{$%
f_{0}(1370)$}$, f_{0}(1710)$ & $89.7\%\ldots 94.1\%$ & $-73.4^{o}\ldots {%
-68.8}$ \\ 
1 $^{3}P_{1}$ & $1^{++}$ & $a_{1}(1260), K_{1A}, $\underline{$f_{1}(1285)$}$%
, f_{1}(1420)$ & $5.4\%\ldots 2.3\%$ & $-41.3^{o}\ldots {-45.9^{o}}$ \\ 
1 $^{3}P_{2}$ & $2^{++}$ & $a_{2}(1320), K_{2}^{*}(1430),$ \underline{$%
f_{2}^{\prime}(1525)$}$, f_{2}(1270)$ & $99.2\%\ldots 99.6\%$ & $%
30.1^{o}\ldots {31.5^{o}}$ \\ 
1 $^{1}D_{2}$ & $2^{-+}$ & $\pi_{2}(1670), K_{2}(1770),$ \underline{$%
\eta_{2}(1870)$}$,\eta_{2}(1645)$ & $99.7\%\ldots 99.8\%$ & $-59.8^{o}\ldots 
{-59.8^{o}}$ \\ 
1 $^{3}D_{3}$ & $3^{--}$ & $\rho_{3}(1690), K_{3}^{*}(1780),$ \underline{$%
\phi_{3}(1850)$}$, \omega_{3}(1670)$ & $99.5\%\ldots 99.8\%$ & $%
31.4^{o}\ldots {32.4^{o}}$ \\ 
1 $^{3}F_{4}$ & $4^{++}$ & $a_{4}(2040), K_{4}^{*}(2045), $\underline{$%
f_{4}(2050)$}$, f_{4}(2220)$ & $0.3\%\ldots 0.2\%$ & $-51.5^{o}\ldots {%
-52.4^{o}}$ \\ 
2 $^{1}S_{0}$ & $0^{-+}$ & $\pi(1300), K(1460), $\underline{$\eta(1440)$}$,
\eta(1295)$ & $\sim 100\%\ldots \sim 100\%$ & $-55.4^{o}\ldots {-55.2^{o}}$
\\ 
2 $^{3}S_{1}$ & $1^{--}$ & $\rho(1300), K^{*}(1410),$ \underline{$%
\omega(1420)$}$, \phi(1680) $ & $10.3\%\ldots 3.9\%$ & $54.0^{o}\ldots
46.7^{o}$ \\ \hline\hline
\end{tabular}
}
\end{center}
\end{table}

\newpage

\begin{table}[tbp]
\caption{Branching ratios and electromagnetic decay widths involving the
axial mesons. $f_{1}$ and $\widetilde{f}_{1}$ stand for $f_{1}(1285)$ and $%
f_{1}(1420)$, respectively. The values presented in our model correspond to
the range $m_{s}/m_{u}=1.3\ldots 1.8$.}
\label{t2}
\begin{center}
\begin{tabular}{|c|c|c|}
\hline\hline
Observable & Our model & Experiment \cite{PDG3} \\ \hline
$\frac{\Gamma _{\gamma \gamma }(\widetilde{f}_{1})}{\Gamma _{\gamma
\gamma}(f_{1})}$ & $0.43\ldots0.29$ & $\frac{\Gamma _{\gamma \gamma }(%
\widetilde{f}_{1})}{\Gamma _{\gamma \gamma}(f_{1})}> \frac{1.4\pm 0.8}{B(%
\widetilde{f}_{1}\rightarrow K\overline{K}\pi )} $ \\ 
$\frac{\Gamma _{\gamma \gamma }(\widetilde{f}_{1})}{\Gamma _{\gamma
\gamma^{*}}(f_{1})}$ & $0.43\ldots0.29$ & $\frac{\Gamma _{\gamma \gamma }(%
\widetilde{f}_{1})}{\Gamma _{\gamma \gamma}(f_{1})}= \frac{0.63\pm 0.34}{B(%
\widetilde{f}_{1}\rightarrow K\overline{K}\pi )} $ \\ 
$\frac{B(J/\psi \rightarrow \gamma \widetilde{f}_{1})}{B(J/\psi
\rightarrow\gamma f_{1})}$ & $1.11\ldots0.81$ & $\frac{1.36\pm 0.44}{B(%
\widetilde{f}_{1}\rightarrow K\overline{K}\pi )}$ \\ 
$\frac{B(f_{1}\rightarrow \gamma \phi )}{B(f_{1}\rightarrow \gamma \rho )}$
& $0.005\ldots0.002$ & $0.013\pm 0.008$ \\ \hline\hline
\end{tabular}
\end{center}
\end{table}

\begin{table}[tbp]
\caption{Branching ratios involving the tensor mesons. $f_{2}$ and $%
\widetilde{f}_{2}$ stand for $f_{2}^{\prime }(1525)$ and $f_{2}(1270)$,
respectively. The values presented in our model correspond to the range $%
m_{s}/m_{u}=1.3\ldots 1.8$.}
\label{t3}
\begin{center}
\begin{tabular}{|c|c|c|}
\hline\hline
Observable & Our model & Experiment \cite{PDG3} \\ \hline
$\frac{B(f_{2}\rightarrow \pi \pi )}{B(f_{2}\rightarrow K\overline{K})}$ & $%
0.024\ldots0.012$ & $0.0092\pm 0.0018$ \\ 
$\frac{B(f_{2}\rightarrow \pi \pi )}{B(f_{2}\rightarrow K\overline{K})}$ & $%
0.18\ldots0.17$ & $0.055_{-0.006}^{+0.005}$ \\ 
$\frac{B(J/\psi \rightarrow \gamma f_{2})}{B(J/\psi \rightarrow \gamma 
\widetilde{f}_{2})}$ & $0.25\ldots0.28$ & $0.34\pm 0.08$ \\ \hline\hline
\end{tabular}
\end{center}
\end{table}

\newpage


\begin{thebibliography}{99}
\bibitem{GMO}  M. Gell-Mann, California Institute of Technology Report
CTSL-20 (1961); S. Okubo, Prog. Theoret. Phys. \textbf{27}, 949 (1962).

\bibitem{schw}  J. Schwinger, Phys. Rev. Lett. \textbf{12}, 273 (1964).

\bibitem{close}  F.E. Close, \textit{An Introduction to Quarks and Partons.}
Academic Press (1979).

\bibitem{ruju}  A. De Rujula, H. Georgi and S.L. Glashow, Phys. Rev. D 
\textbf{12}, 147 (1975).

\bibitem{w1}  W.S. Carvalho, A.C.B. Antunes and A.S. de Castro, Mod. Phys.
Lett. A\textbf{\ 12}, 121 (1997).

\bibitem{PDG1}  Particle Data Group, L. Montanet \textit{et al.}, Phys. Rev.
D \textbf{50}, 1173 (1994).

\bibitem{w2}  W.S. Carvalho, A.C.B. Antunes and A.S. de Castro, Hadronic J. 
\textbf{22}, 105 (1999).

\bibitem{scharre}  D.L. Scharre \textit{et al.}, Phys. Lett. B \textbf{97},
329 (1980); C. Edwards \textit{et al.}, Phys. Rev. Lett.\textbf{49}, 259
(1982).

\bibitem{japa}  N. Aizawa, Z. Maki and I. Umemura, Prog. Theor. Phys. 68,
2120 (1982).

\bibitem{rosner}  J.L. Rosner, Phys. Rev. D \textbf{27}, 1101 (1983); J.L.
Rosner and S.F. Tuan, Phys. Rev. D \textbf{27} 1544 (1983).

\bibitem{balt}  R.M. Baltrusaitis \textit{et al.}, Phys. Rev. D \textbf{32},
2883 (1985).

\bibitem{antunes}  F. Caruso, E. Predazzi, A.C.B. Antunes and J. Tiommo, Z.
Phys. C \textbf{30}, 493 (1986).

\bibitem{jousset}  J. Jousset \textit{et al.}, Phys. Rev. D \textbf{41},
1389 (1990).

\bibitem{bai}  Z. Bai \textit{et al.}, Phys. Rev. Lett. \textbf{65}, 2057
(1990); C. Amsler \textit{et al.}, Phys. Lett. B \textbf{358}, 389 (1995);
A. Bertin \textit{et al.}, Phys. Lett. B \textbf{361}, 187 (1995).

\bibitem{kitamura}  I. Kitamura, N. Morisita and T. Teshima, Int. J. Mod.
Phys. A \textbf{31}, 5489 (1994).

\bibitem{closea}  F.E. Close, G.R. Farrar and Z. Li, Phys. Rev. D \textbf{55}%
, 5749 (1997).

\bibitem{pred}  M. Genovese, D.B. Lichtenberg and E. Predazzi, Z. Phys. C 
\textbf{61}, 425 (1994).

\bibitem{w3}  W.S. Carvalho, A.C.B. Antunes and A.S. de Castro, Eur. Phys.
J. C \textbf{7}, 95 (1999).

\bibitem{lipkinb}  H.J. Lipkin, Phys. Lett. B \textbf{67}, 65 (1977).

\bibitem{PDG2}  Particle Data Group, R.M. Barnett \textit{et al.}, Phys.
Rev. D \textbf{54}, 1 (1996).

\bibitem{k1}  G.M. Brandenburger \textit{et al., }Phys. Rev. Lett. \textbf{36%
}, 703 (1976); R.K. Carnegie \textit{et al.}, Nucl. Phys. B \textbf{127},
509 (1977); M.G. Bowler, J. Phys. G \textbf{3}, 775 (1977).

\bibitem{Gavillet}  P. Gavillet \textit{et al.}, Z. Phys. C. \textbf{16},
119 (1982); D. Aston \textit{et al.}, Phys. Lett. B \textbf{201}, 573 (1988).

\bibitem{Bityukov}  S.I. Bityukov \textit{et al.,} Phys. Lett B \textbf{203}%
, 327 (1988); S. Ishida \textit{et al.,} Prog. Theor. Phys. \textbf{82}, 119
(1989).

\bibitem{Alde}  D. Alde \textit{et al.,} Phys. Lett. B \textbf{241,} 600
(1990); D.V. Bugg \textit{et al.,} Phys. Lett. B \textbf{353}, 378 (1995).

\bibitem{Augustin}  J.E. Augustin \textit{et al.,} Phys. Rev. Lett. \textbf{%
60, }2238 (1988); T.A. Armstrong \textit{et al.,} Phys. Lett. B \textbf{227}%
, 186 (1989).

\bibitem{bes}  J.Z. Bai \textit{et al.}, Phys. Rev. Lett. \textbf{77}, 3959
(1996).

\bibitem{w4}  W.S. Carvalho, A.S. de Castro and A.C.B. Antunes, Eur. Phys.
J. C \textbf{17}, 173 (2000).

\bibitem{PDG3}  Particle Data Group, D.E. Groom \textit{et al}., Eur. Phys.
J. C \textbf{15}, 1 (2000).

\bibitem{harber}  H.E. Haber and J. Perrier, Phys. Rev. D \textbf{32}, 2961
(1985); I. Bediaga, F. Caruso and E. Predazzi, Nuovo Cim. A \textbf{91}, 306
(1986); F. Caruso and E. Predazzi, Europhys. Lett. \textbf{6}, 677 (1987);
A. Bramon and M.D. Scadron, Phys. Lett. B \textbf{234}, 346 (1990); C.
Amsler \textit{et al.}, Phys. Lett. B \textbf{194}, 451 (1992); P. Ball 
\textit{et al.}, Phys. Lett. B \textbf{365}, 367 (1996); M. Genovese,
hep-ph/9608451; G.R. Farrar, hep-ph/96123547; F.E. Close, Nucl. Phys. Proc.
Suppl. A \textbf{56}, 248 (1997); A.V. Anisovich, V.V. Anisovich and A.V.
Sarantsev, Phys. Lett. B \textbf{395}, 123 (1997); A. Bramon \textit{et al.}%
, Eur. Phys. J. C \textbf{7}, 271 (1999); L. Burakovsky and T. Goldman,
Phys. Rev. D \textbf{57}, 2879 (1998).

\bibitem{fclose}  F.E. Close and A. Kirk, Z. Phys. C \textbf{76}, 469 (1997).

\bibitem{fritzch1}  H. Fritzch and P. Minkowsky, Nuovo Cim. A \textbf{30},
393 (1975); H. Fritzch and J.D. Jackson, Phys. Lett. B \textbf{66}, 365
(1977); N. Isgur, Phys. Rev. D \textbf{21}, 779 (1980); H.J. Schnitzer,
Nucl. Phys. B \textbf{207}, 131 (1982); T. Teshima and S. Oneda, Phys. Rev.
D \textbf{27}, 1551 (1983); S. Godfrey and N. Isgur, Phys. Rev. D \textbf{34}%
, 899 (1986); F.J. Gilman and R. Kauffman, Phys. Rev. D \textbf{36}, 2761
(1987); T. Teshima, I. Kitamura and N. Morisita, Nuovo Cim. A \textbf{103},
175 (1990); M. Birkel and H. Fritsch, Phys. Rev. D \textbf{53}, 6195 (1996);
M.M. Brisudova \textit{et al.}, Phys. Rev. D \textbf{58}, 114015 (1998); D.
Weingarten, Nucl. Phys. (Proc. Suppl.) B \textbf{53}, 232 (1997); H.-M. Choi
and C.-R. Ji, Phys. Rev. D \textbf{59}, 074015 (1999); L. Burakovsky and T.
Goldman, Nucl. Phys. A \textbf{628}, 87 (1998).

\bibitem{lipkina}  I. Cohen and H.J. Lipkin, Nucl. Phys. B \textbf{151}, 16
(1979).

\bibitem{isgur2}  N. Isgur, Phys. Rev. D \textbf{21}, 779 (1980).

\bibitem{li}  D.-M. Li, H. Yu and Q.-X. Shen, Chin. Phys. Lett. \textbf{17},
558 (2000).

\bibitem{li2}  D.-M. Li, H. Yu and Q.-X. Shen, J. Phys. G \textbf{27}, 807
(2001).

\bibitem{feclose}  F.E. Close and P.R. Page, Nucl. Phys.\textbf{\ 443}, 233
(1995).

\bibitem{ado}  J. Adomeit \textit{et al.}, Z. Phys. C \textbf{71}, 227
(1996).

\bibitem{go}  S. Godfrey, R. Kokoski and N. Isgur, Phys. Lett. B \textbf{141}%
, 439 (1984).

\bibitem{blu}  H. Blundell and S. Godfrey, Phys. Rev. D \textbf{53}, 3700
(1996).

\bibitem{asei}  A. Seiden, H.F.-W. Sadrozinski and H.E. Harber, Phys. Rev. D 
\textbf{38}, 824 (1988).

\bibitem{lass}  D. Aston \textit{et al}., Nucl. Phys. B (Proc. Suppl.) 
\textbf{21}, 105 (1991).
\end{thebibliography}
\end{document}